\documentclass[a4paper]{jpconf}

\usepackage{hyperref}
\usepackage{graphicx}
\usepackage{amssymb}
\usepackage{latexsym}

\begin{document}

\title{Is there potential complementarity between LISA and pulsar timing?}

\author{Matthew Pitkin, James Clark, Martin A Hendry, Ik Siong Heng, Chris
Messenger, Jennifer Toher, Graham Woan}
\address{Department of Physics and Astronomy, Kelvin Building,
University of Glasgow, Glasgow G12 8QQ, UK}
\ead{matthew@astro.gla.ac.uk}

\begin{abstract}
We open the discussion into how the Laser Interferometer Space Antenna (LISA)
observations of supermassive black-hole (SMBH) mergers (in the mass range
$\sim10^6-10^8\,{\rm M}_{\odot}$) may be complementary to pulsar timing-based
gravitational wave searches. We consider the toy model of determining pulsar
distances by exploiting the fact that LISA SMBH inspiral observations can place
tight parameter constraints on the signal present in pulsar timing observations.
We also suggest, as a future path of research, the use of LISA ring-down
observations from the most massive ($\gtrsim {\rm a ~few~} 10^7\,{\rm
M}_{\odot}$) black-hole mergers, for which the inspiral stage will lie outside
the LISA band, as both a trigger and constraint on searches within pulsar timing
data for the inspiral stage of the merger.
\end{abstract}

\section{Introduction}
A major source of strong gravitational wave (GW) signals for the future
space-based GW detector LISA \cite{LISAjpl, LISAesa} will be the inspiral and
ring-down of merging supermassive black-holes (SMBHs) at cosmological distances
(e.g \cite{Hughes:2002}.) The primary sensitive frequency band for LISA is
0.0001--0.1\,Hz and the most abundant SMBH mergers in this range will have
component masses in the $10^4-10^6\,{\rm M}_{\odot}$ range. Inspirals can
potentially be seen for SMBH systems with masses up to a few $10^7\,{\rm
M}_{\odot}$, although at greater masses than this the frequency at which they
reach their last stable orbit and merge will be outside the sensitive LISA band
(i.e. $\lesssim 10^{-4}$\,Hz.) More massive systems up to a few $10^8\,{\rm
M}_{\odot}$ merge before they enter the LISA frequency band, but are observable
in their ring-down phase \cite{Hughes:2002}, with characteristic frequency given by
$f_c\approx1.3\times10^{-3}(10^7\,{\rm M}_{\odot}/M)$\,Hz (where $M$ is the
post-coalescence black-hole mass.) Such systems will potentially be observed at
signal-to-noise ratios of hundreds to thousands allowing their waveform to be
precisely parameterised. Estimates of event rates vary greatly between authors
(see \cite{Berti:2006} for a summary of estimates), but could range from tens to
thousands of events over a range of distances and signal-to-noise ratios. 

With pulsar timing there is another method to observe low frequency GWs. This
requires stable millisecond pulsars with low intrinsic noise and very accurate
pulsar timing models \cite{Hobbs:2006, Edwards:2006} to produce timing
residuals that could contain a GW signal. Current experiments (e.g. the Parkes
Pulsar Timing Array - PPTA \cite{Manchester:2006}), with sample rates of order
two weeks or so, are sensitive to GWs with frequencies $\lesssim 10^{-6}$\,Hz.
However, in the future with the Square Kilometre Array (SKA) daily sampling may
be possible giving an upper limit on frequencies of $\lesssim 10^{-5}$\,Hz. For
the higher mass SMBH systems ($\gtrsim 10^{7}\,{\rm M}_{\odot}$) these
frequencies would have been swept through thousands of years prior to the final
stage inspiral observed by LISA, but pulsars, at distances of several
kiloparsecs and therefore affected by the past GW signal from when the
pulsar's pulses were emitted, may contain signals at these low frequencies. This
suggests that there may be some overlap or complementarity between the two types
of observation which could be exploited to gain the maximum astrophysical
information. 

The ideas presented here can be compared to previous work \cite{Lommen:2001,
Jenet:2004} in which upper limits on potential nearby SMBH binaries were placed
using pulsar timing. In those studies assumptions about the sources, based on
some, maybe speculative, observational evidence were used to constrain the
system models and produce upper limits on the system parameters. Here we suggest
using LISA observations as the constraint on the models with which we search in
pulsar timing data. The non-observation of inspirals in both LISA and pulsar
timing from known recently merged galaxies would place limits on SMBH systems at
different stages of the binary evolution or with different mass ranges.

\section{Pulsar timing}
Many pulsars are very precise clocks. The residuals in their timing, after the
removal of an observationally-fitted model of the pulsar's phase evolution, are
a potential probe of the low frequency GW spectrum \cite{Detweiler:1979}.
Residuals contain noise from signal processing and the intrinsic instability of
the pulsar. The PPTA \cite{Manchester:2006} aims to time tens of the most stable
pulsars with precisions of around 100\,ns. In the future with the SKA it may be
possible to time thousands of millisecond pulsars, some of which with a
precision of $<100$\,ns \cite{SKABook}. There have already been efforts to
search for a GW background \cite{Jenet:2006}, and individual systems
\cite{Jenet:2004}, of SMBH inspirals in existing pulsar timing data.

\begin{figure}
\begin{center}
\includegraphics[width=0.55\textwidth]{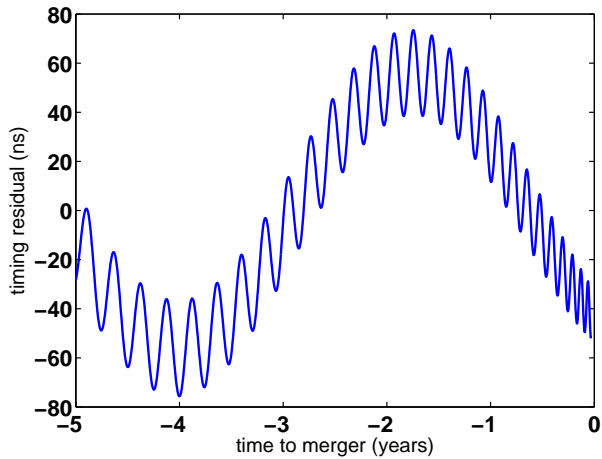}
\caption{The residual left in the timing of a pulsar at a distance of
$\sim3$\,kpc, with an angular separation between the source and the pulsar of
$\pi/2$\,rads, caused by the coalescence of two $5\times10^9\,{\rm M}_{\odot}$
black holes at a distance of 1\,Gpc.}
\label{fig:exampleresidual}
\end{center}
\end{figure}

Pulsar timing residuals contain two components of a gravitational wave signal:
the part passing the Earth (which will be correlated between all pulsar
observations); and the part passing the pulsar as it emitted the pulses now
being observed. The residual amplitude from the signal will depend on (other 
than the intrinsic GW strain) the angular separation between the GW source and
the pulsar, with sources along the pulsar line-of-sight producing no residual.
The timing residual amplitude will also increase with the source period. An
example of the pre-fit timing residuals that would remain in the timing of a
perfectly modeled pulsar $\sim3$\,kpc away, due to the inspiral of two
$5\times10^9\,{\rm M}_{\odot}$ SMBHs\footnote{This system would be outside the
LISA band, but is used for ease of visualisation.}, can be seen in
Figure~\ref{fig:exampleresidual}.
The low frequency component of the residual from the signal passing the pulsar
can easily be distinguished from the high frequency component as the signal
passes the Earth. Pulsar timing residuals observed before the SMBH merger is
seen in LISA would contain both components of the signal, although the high
frequency component would likely be too high to be observed, but pulsar
observations after the event is seen in LISA would only contain the low
frequency (i.e. at the pulsar) component.

LISA is expected to be launched within the next 10--15 years, with a similar
timescale to the proposed development of the SKA, meaning there could be
overlap between their operation. On these timescales the pulsar community hope
to be timing many pulsars with $\sim100$\,ns (or less) precision. However,
overlap between these two experiments is not necessary for complementarity to
exist between LISA and pulsar timing, as archive pulsar timing data could be
looked at once LISA observations have been made, or vice versa. Thus even
current pulsar timing data sets could be useful, and would indeed give a longer
time baseline of data to study in the future. 

\section{Pulsar distance measurements}
First we will discuss a toy model showing one way in which LISA observations and
pulsar timing could be complementary under a set of extremely optimistic
assumptions: a method of pulsar distance determination. This application is more
for use as a way of opening up discussion into other areas of complementarity
than for its real world applicability. 

For the high signal-to-noise ratio inspirals seen with LISA the parameters
describing the signal, including the source sky location, can be very well
constrained. Consequently the equivalent waveform present in the pulsar timing
data can be approximated such that the only unknown is the distance to the
pulsar. This provides a way of obtaining pulsar distances, for the stable
millisecond pulsars used in GW searches at least, independently of the galactic
electron density distribution model \cite{Taylor:1993} used in dispersion
measure distance estimates, which can have large uncertainties
\cite{Frail:1990}. However, for such pulsars high precision measurements of
parallax or the orbital period derivative have, and will in the future, be used
to give extremely precise distance measurements with errors of order
$\sim\pm1\%$ (see e.g. \cite{Verbiest:2008, SKAbookb}.) Here we will see if our
method can be competitive with these other methods.

\begin{figure}
\begin{center}
\includegraphics[width=0.7\textwidth]{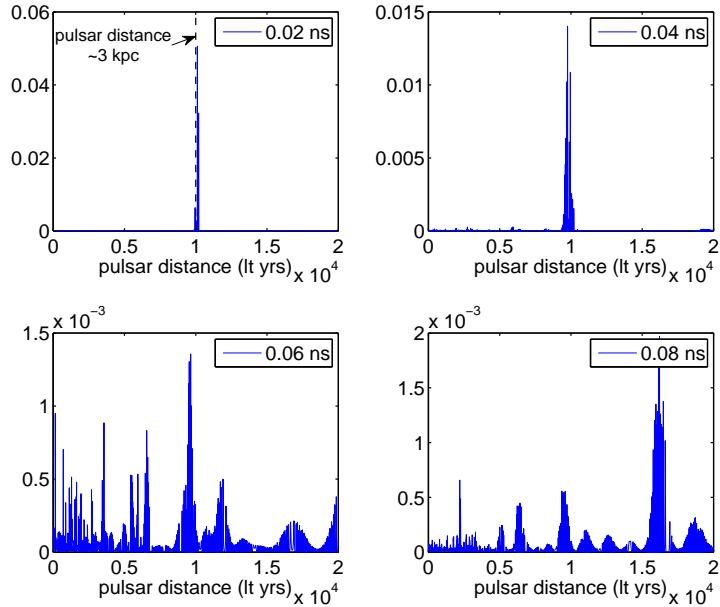}
\caption{The probability distribution of the pulsar distance calculated using
simulated timing residual data containing an injected inspiral signal from two
$5\times10^7\,{\rm M}_{\odot}$ black holes at $z=1$ and Gaussian white noise
with standard deviations of 0.02, 0.04, 0.06 and 0.08\,ns.}
\label{fig:distanceestimate}
\end{center}
\end{figure}

We create five years of data prior to a LISA observed merger, consisting of
500 equally spaced observations, containing a signal from a simulated system of
two black holes with redshifted masses of $5\times10^7\,{\rm M}_{\odot}$ at
$z=1$. This kind of system would be just within the LISA sensitive band before
merger. We consider a pulsar with an angular separation of $45^{\circ}$ from the
SMBH binary, at a distance of $\sim3$\,kpc, and, to estimate the limiting
potential of this method, firstly subject to a highly \emph{un}realistic timing
residual noise\footnote{This residual noise is Gaussian and white and not
representative of real world residuals in either its amplitude or spectral
shape.} of around 0.02\,ns (note that expected residuals will be of order
10--100\,ns.) In this case the pulsar distance can be determined to about 1--2\%
accuracy (see Figure~\ref{fig:distanceestimate}.) However, if the residual noise
becomes just a few times larger the distance estimation will substantially
degrade.
To obtain pulsar distances with a few percent error (e.g. comparable to those
in \cite{Verbiest:2008}), using the potentially obtainable timing residuals of
100\,ns (with the PPTA) up to 10\,ns \cite{Bailes:2007} would require the SMBH
system to be $\sim0.5$ and $\sim5$\,Mpc away respectively. Such high mass
($\gtrsim 5\times10^7\,{\rm M}_{\odot}$) inspirals within these distance
ranges are extremely rare (see \cite{Wyithe:2003} for predictions of event
rates) and lower mass systems would need to be even closer to give decent pulsar
distance estimates. From this study it can be seen that the orbital period
derivative and parallax methods show far better promise as a realisable and
accurate pulsar distance measurement than our method.

\section{Ring-down signals as triggers}
The ring-down of the SMBH formed after mergers offers a better LISA detection
rate for high mass systems (greater than a few $10^7\,{\rm M}_{\odot}$), due to
their higher signal frequency and visibility to larger distances
\cite{Hughes:2002}. However, LISA will be unable to obtain information about the
inspiral phase for these systems. Pulsar timing observations might, then,
provide an opportunity to probe this unseen inspiral stage of the merger. Some
constraints on the system parameter space and the time of coalescence seen by
LISA will yield a trigger with which to search in pulsar timing data. Current
pulsar distance estimates could be used to aid the search. Unfortunately the
ring-down does not provide information on the source's sky position making the
parameter space more complicated, although there may be other electromagnetic
counterparts to the signal that could give the source position. Preliminary work
suggests that to see any of the SMBH ring-downs that occur at a reasonable rate
within the LISA band (i.e. are at cosmological distances), would be impossible
with any single pulsar observations using current and projected achievable
timing residuals. However, if we consider the fact that observations of multiple
pulsars will be made (potentially hundreds to thousands of separate objects with
the SKA) then a global fit using data from them all will help dig into the noise
and could reveal a signal. Multiple pulsar observations could also be used to
estimate the source position. This work will be discussed more thoroughly in a
future paper.

\section{Conclusions}
With the advent of dedicated pulsar timing of the most stable pulsars direct
GW detection via measurements of their timing residuals may soon be possible.
There are certain events which could overlap between being observed in the
planned space-based GW detector LISA (or even more distant future space-based
detectors) and the pulsar timing data. We have suggested some ways in which
such measurements could be complementary and reveal astrophysics that would be
more difficult, or unobtainable, with one alone. Using current, and near term
projected, pulsar timing accuracies, the measurement of pulsar distances as
described above would require a very serendipitous event. The reconstruction
of the inspiral, for events only observed as ring-downs in LISA, again would
likely require more very stable pulsars observations and a serendipitous loud,
nearby merger. This area of investigation is intended for further study and
will hopefully reveal other areas of complementarity.

\section*{Acknowledgements}
The authors would like to thank the organisers of the Amaldi conference for an
excellent meeting. The authors are also very grateful to George Hobbs for the
many constructive and useful comments during the preparation of this manuscript.


\end{document}